\documentclass[aps,prd,twocolumn,superscriptaddress]{revtex4}
\usepackage{epsfig,epsf}
\usepackage{amsmath}
\usepackage{amsthm}
\usepackage{amsfonts}
\usepackage{amssymb}
\usepackage{dsfont}
\usepackage{multirow}
\usepackage{appendix}
\usepackage{slashed}
\usepackage[active]{srcltx}
\usepackage{psfrag}

\begin{document}

\title{Light axial-vector and vector resonances $X(2100)$ and $X(2239)$}
\author{K.~Azizi}
\affiliation{Department of Physics, University of Tehran, North Karegar Ave., Tehran
14395-547, Iran}
\affiliation{Department of Physics, Do\v{g}u\c{s} University, Acibadem-Kadik\"{o}y, 34722
Istanbul, Turkey}
\author{S.~S.~Agaev}
\affiliation{Institute for Physical Problems, Baku State University, Az--1148 Baku,
Azerbaijan}
\author{H.~Sundu}
\affiliation{Department of Physics, Kocaeli University, 41380 Izmit, Turkey}

\begin{abstract}
We study features of the resonances $X(2100)$ and $X(2239)$ by treating them
as the axial-vector and vector tetraquarks with the quark content $ss%
\overline{s}\overline{s}$, respectively. The spectroscopic parameters of
these exotic mesons are calculated in the framework of the QCD two-point sum
rule method. Obtained prediction for the mass $m=(2067 \pm 84)~\mathrm{MeV}$
of the axial-vector state is in excellent agreement with the mass of the
structure $X(2100)$ recently observed by the BESIII Collaboration in the
decay $J/\psi \to \phi \eta \eta^{\prime}$ as the resonance in the $\phi
\eta^{\prime}$ mass spectrum. We explore also the $S$-wave decays $X(2100)
\to \phi\eta^{\prime}$ and $X(2100) \to \phi \eta$ using the QCD light-cone
sum rule approach and technical methods of the soft-meson approximation. The
width of the axial-vector tetraquark, $\Gamma=(130.2\pm 30.1)~\mathrm{MeV}$,
saturated by these two decays is comparable with the measured full width of
the resonance $X(2100)$. Our prediction for the vector $ss\overline{s}%
\overline{s}$ tetraquark's mass $\widetilde{m}=(2283\pm 114)~\mathrm{MeV}$
is consistent with the experimental result $2239.2 \pm 7.1 \pm 11.3~\mathrm{%
MeV}$ of the BESIII Collaboration for the mass of the resonance $X(2239)$.
\end{abstract}

\maketitle


\section{Introduction}

\label{sec:Int}
Hadrons with exotic structures and/or quantum numbers, which differ them
from the conventional $\bar{q}q$ mesons and $qq^{\prime }q^{\prime \prime }$
baryons were and remain in agenda of the High Energy Physics community.
Properties of the ordinary hadrons, i.e. their spectroscopic parameters as
well as their strong, semileptonic and radiative transitions have been
investigated in the framework of Quantum Chromodynamics (QCD), and
successfully confronted with available experimental data. In the
nonperturbative regime of momentum transfers, the relevant theoretical
results have been obtained using methods and phenomenological models which
use either the first principles of QCD or invoke additional assumptions
about the internal structure and dynamics of hadrons.

At the same time, the QCD allows existence of not only the ordinary hadrons
but also particles built of four, five, or more quarks, quark-gluon hybrids,
and glueballs. The idea about the multi-quark nature of some observed
particles was first applied to explain the unusual features of the light
scalar mesons with masses $m<1~\mathrm{GeV}$ \cite{Jaffe:1976ig}. The reason
is that the nonet of scalar particles in the standard model of mesons should
be realized as $1{}^{3}P_{0}$ quark-antiquark states. But masses of these
scalars, in accordance with various model computations, are higher than $1~%
\mathrm{GeV}$. Moreover, the standard model could not correctly describe the
mass hierarchy of the mesons inside the nonet. These problems can be evaded
by assuming that the light scalars are four-quark exotic mesons, or at least
contain substantial four-quark component. In the context of this scheme low
masses of the scalar mesons, as well as the hierarchy inside of the nonet
receive natural explanations. A recent model of the both light and heavy
scalar nonets is based on suggestion about diquark-antidiquark structure of
these particles which are mixtures of the spin-$0$ diquarks from ($\overline{%
\mathbf{3}}_{c},\overline{\mathbf{3}}_{f}$) representation with spin-$1$
diquarks from ($\mathbf{6}_{c},\ \overline{\mathbf{3}}_{f})$ representation
of the color-flavor group \cite{Kim:2017yvd}. The spectroscopic parameters
and width of the light scalar mesons $f_{0}(500)$ and $f_{0}(980)$
calculated by considering them as admixtures of the $SU_{f}(3)$ flavor octet
and singlet tetraquarks are in a reasonable agreement with experimental data
\cite{Agaev:2017cfz,Agaev:2018sco}. Other members of the light scalar nonet
were also successfully explained as scalar particles with relevant
diquark-antidiquark contents \cite{Agaev:2018fvz}.

However, light quarks may not form stable tetraquarks: Theoretical studies
proved that only tetraquarks composed of heavy and light diquarks may be
stable against the strong decays. Thus, four-quark systems $QQ\bar{Q}\bar{Q}$
and $QQ\bar{q}\bar{q}$ were studied in Refs.\ \cite%
{Ader:1981db,Lipkin:1986dw,Zouzou:1986qh} by employing the conventional
potential model with additive pairwise interaction of color-octet exchange
type. Within this approach it was demonstrated that states $QQ\bar{q}\bar{q}$
may form the stable composites provided that the ratio $m_{Q}/m_{q}$ is
large enough. Experimental information on possible tetraquark candidates is
also connected with the heavy resonances observed in various processes.
Starting from discovery of the charmonium-like resonance $X(3872)$ by Belle
Collaboration \cite{Choi:2003ue}, the exotic mesons are the objects of
rapidly growing studies. Valuable experimental data collected during years
passed from observation of the $X(3872)$ resonance, as well as important
theoretical achievements form now the physics of the exotic hadrons \cite%
{Chen:2016qju,Chen:2016spr,Esposito:2016noz,Ali:2017jda,Olsen:2017bmm}.

There are only few resonances seen in the experiments which may be
considered as four-quark systems containing only the light quarks. One of
such states is the famous structure $Y(2175)$ discovered by the BaBar
Collaboration in the process $e^{+}e^{-}\rightarrow \gamma _{\mathrm{ISR}%
}\phi f_{0}(980)$ as a resonance in the $\phi f_{0}(980)$ invariant mass
spectrum \cite{Aubert:2006bu}. Existence of the $Y(2175)$ later was
confirmed by the BESII, Belle, and BESIII collaborations as well \cite%
{Ablikim:2007ab,Shen:2009zze,Ablikim:2014pfc}. The mass and width of this
state with spin-parities $J^{PC}=1^{--}$ is $m=2175 \pm 10 \pm 15~\mathrm{MeV%
}$ and $\Gamma=58 \pm 16 \pm 20~\mathrm{MeV}$, respectively.

Other resonances which may be interpreted as light exotic mesons were
observed recently by the BESIII Collaboration. Thus, the $X(2239)$ was seen
in the process $e^{+}e^{-}\rightarrow K^{+}K^{-}$ as a resonant structure in
the cross section shape line \cite{Ablikim:2018iyx}. The mass and width of
this state were found equal to $m=2239.2\pm 7.1\pm 11.3~\mathrm{MeV}$ and $%
\Gamma =139.8\pm 12.3\pm 20.6~\mathrm{MeV}$, respectively. The $X(2100)$ was
fixed in the process $J/\psi \rightarrow \phi \eta \eta ^{\prime }$ as a
resonance in the $\phi \eta ^{\prime }$ mass spectrum \cite{Ablikim:2018xuz}%
. The collaboration studied the angular distribution of $J/\psi \rightarrow
X(2100)\eta $ , but due to limited statistics could not clearly distinguish $%
1^{+}$ or $1^{-}$ assumption for the spin-parity $J^{P}$ of the $X(2100)$. \
Therefore, the spectroscopic parameters of this resonance were determined
using both of these assumptions. In the case $J^{P}=1^{-}$ \ the mass and
width of the $X(2100)$ were measured to be $m=2002.1\pm 27.5\pm 21.4~\mathrm{%
MeV}$ and $\Gamma =129\pm 17\pm 9~\mathrm{MeV}$. Alternatively, the
assumption $J^{P}=1^{+}$ led to the results $m=2062.8\pm 13.1\pm 7.2~\mathrm{%
MeV}$ and $\Gamma =177\pm 36\pm 35~\mathrm{MeV}$.

Theoretical interpretations of these light resonances which may be
considered as candidates for tetraquarks, as usua,l comprise all possible
models and approaches available in high energy physics. Because the $Y(2175)$
was discovered more than ten years ago, there are numerous and diverse
articles in the literature devoted to its investigation. There are quite
natural attempts to interpret it as an $2{}^{3}D_{1}$ excitation of the
conventional $\overline{s}s$ meson \cite{Ding:2007pc,Wang:2012wa}. Another
traditional approach is to treat such states as dynamically generated
resonances. As a dynamically generated state in the $\phi K\overline{K}$
system, the $Y(2175)$ was examined in Ref.\ \cite{MartinezTorres:2008gy}.
The similar dynamical picture may appear due to self-interaction between $%
\phi $ and $f_{0}(980)$ mesons as well \cite{AlvarezRuso:2009xn}.
Alternative explanations of the $Y(2175)$ resonance's structure include a
hybrid meson $\overline{s}sg$, or a baryon-antibaryon $qqs\overline{q}%
\overline{q}\overline{s}$ state that couples strongly to the $\Lambda
\overline{\Lambda } $ channel (for relevant references and other models, see
Ref.\ \cite{Ablikim:2018iyx}).

The resonance $Y(2175)$ as a vector tetraquark with $s\overline{s}s\overline{%
s}$ or $ss\overline{s}\overline{s}$ content was explored in Refs.\ \cite%
{Wang:2006ri} and \cite{Chen:2008ej,Chen:2018kuu}, respectively. In these
works the authors used the QCD sum rule method and evaluated spectroscopic
parameters of these states. The newly found structures $X(2100)$ and $%
X(2239) $ (hereafter $X_{1}$ and $X_{2}$, respectively) were also analyzed
as vector or axial-vector tetraquarks. Thus, in Ref.\ \cite{Lu:2019ira} the
mass spectrum of the $ss\overline{s}\overline{s}$ tetraquark states was
investigated within the relativized quark model. The authors concluded that
the resonance $X_{2}$ can be assigned as a $P$-wave $1^{--}$ $ss\overline{s}%
\overline{s}$ tetraquark. In the framework of the QCD sum rule method the $%
X_{1}$ resonance was studied in Refs.\ \cite{Cui:2019roq,Wang:2019nln}.
Predictions obtained there allowed the authors to interpret it as the
axial-vector $ss\overline{s}\overline{s}$ tetraquark with the quantum
numbers $J^{PC}=1^{+-}$. In accordance with Ref.\ \cite{Wang:2019qyy}, the $%
X_{1}$ may be identified as the second radial excitation of the conventional
meson $h_{1}(1380)$.

As is seen, theoretical interpretations of observed light resonances are
numerous and sometimes contradict to each other. There is a necessity to
consider this problem in a more detailed form and analyze not only
spectroscopic parameters of the light resonances, but also to explore their
decay channels and widths. In the present work we study the axial-vector and
vector light tetraquarks $ss\overline{s}\overline{s}$ and compute their
masses and couplings. By confronting theoretical predictions and
experimental data we identify the observed resonances $Y(2175)$, $X_{1}$ and
$X_{2}$ with these tetraquark structures. It turns out that the resonance $%
X_{1}$ can be interpreted as a axial-vector tetraquark state. We calculate
the width of the decays $X_{1}\rightarrow \phi \eta ^{\prime }$ and $%
X_{1}\rightarrow \phi \eta $ which are essential for our interpretation of
the $X_{1}$. Among the vector resonances $Y(2175)$ and $X_{2}$, parameters
of the latter is closer to our result.

Calculations in the present paper are performed in the context of the QCD
sum rule method, which is one of the powerful nonperturbative approaches in
high energy physics \cite{Shifman1,Shifman2}. The masses and couplings of
the four-quark systems are evaluated using two-point QCD sum rules with an
accuracy higher than in existing samples. To find the width of the decays $%
X_{1}\rightarrow \phi \eta ^{\prime }$ and $X_{1}\rightarrow \phi \eta $ we
employ sum rules on the light cone and technical tools of the soft-meson
approximation \cite{Balitsky:1989ry,Belyaev:1994zk}.

This paper is structured in the following form: In Sections \ref{sec:Mass1}
and \ref{sec:Mass2} we analyze the spectroscopic parameters of the
axial-vector and vector tetraquarks $ss\overline{s}\overline{s}$ and provide
details of relevant sum rule calculations. In Sec.\ \ref{sec:Decay} the
strong couplings $g_{X_{1}\phi \eta ^{\prime }}$ and $g_{X_{1}\phi \eta }$
corresponding to the vertices $X_{1}\phi \eta ^{\prime }$ and $X_{1}\phi
\eta $ are found using the QCD light-cone sum rule method. These coupling
are required to evaluate the width of the decays $X_{1}\rightarrow \phi \eta
^{\prime }$ and $X_{1}\rightarrow \phi \eta $, respectively. Section \ref%
{sec:Conc} contains summary of the obtained results and our conclusions.


\section{Mass and coupling of the axial-vector tetraquark $ss\overline{s}%
\overline{s}$}

\label{sec:Mass1}
In this section we compute the mass and coupling of the axial-vector
tetraquark $T_{\mathrm{AV}}=ss\overline{s}\overline{s}$. As it has been
emphasized above, to this end we use the QCD sum rules method which is based
on first principles of QCD and allows one, via a quark-hadron duality
assumption, to express physical parameters of hadrons in terms of the
universal nonperturbative quantities, i.e. vacuum expectation values of
local quark, gluon, and mixed operators. This method was successfully
applied to explore parameters not only of conventional hadrons, but also to
study various multi-quark systems \cite{Albuquerque:2018jkn}.

To derive the required sum rules we consider the two-point correlation
function $\Pi _{\mu \nu }(p)$, which is defined by the formula
\begin{equation}
\Pi _{\mu \nu }(p)=i\int d^{4}xe^{ipx}\langle 0|\mathcal{T}\{J_{\mu
}(x)J_{\nu }^{\dag }(0)\}|0\rangle ,  \label{eq:CF1}
\end{equation}%
where $J_{\mu }(x)$ is the interpolating current for the axial-vector
tetraquark $ss\overline{s}\overline{s}$. The choice of $J_{\mu }(x)$ in one
of the main operations in the sum rule computations. The tetraquark with
content $ss\overline{s}\overline{s}$ and spin-parities $J^{PC}=1^{+-}$ can
be interpolated using different currents. The current that leads to a
reliable prediction for the mass and coupling of the axial-vector state has
the following form \cite{Cui:2019roq}
\begin{eqnarray}
&&J_{\mu }(x)=\left[ s_{a}^{T}(x)C\gamma ^{\nu }s_{b}(x)\right] \left[
\overline{s}_{a}(x)\sigma _{\mu \nu }\gamma _{5}C\overline{s}_{b}^{T}(x)%
\right]  \notag \\
&&-\left[ s_{a}^{T}(x)C\sigma _{\mu \nu }\gamma _{5}s_{b}(x)\right] \left[
\overline{s}_{a}(x)\gamma ^{\nu }C\overline{s}_{b}^{T}(x)\right] .
\label{eq:Curr1}
\end{eqnarray}%
Here $a$ and $b$ are the color indices and $C$ is the charge conjugation
operator.

The sum rules necessary to calculate the mass $m$ and coupling $f$ of the $%
T_{\mathrm{AV}}$ can be derived in accordance with prescriptions of the
method, which require first to express the correlation function $\Pi _{\mu
\nu }(p)$ using the tetraquark's physical parameters . We consider $T_{%
\mathrm{AV}}$ as a ground-state particle, and after isolating the first term
in $\Pi _{\mu \nu }^{\mathrm{Phys}}(p)$ get
\begin{equation}
\Pi _{\mu \nu }^{\mathrm{Phys}}(p)=\frac{\langle 0|J_{\mu }|T_{\mathrm{AV}%
}(p)\rangle \langle T_{\mathrm{AV}}(p)|J_{\nu }^{\dagger }|0\rangle }{%
m^{2}-p^{2}}+\dots  \label{eq:CF2}
\end{equation}%
Equation (\ref{eq:CF2}) is obtained by saturating the correlation function
with a complete set of $J^{P}=1^{+-}$ states and carrying out the
integration over $x$. Effects of higher resonances and continuum states are
denoted above by dots.

To simplify further the correlator $\Pi _{\mu \nu }^{\mathrm{Phys}}(p)$, it
is convenient to introduce the matrix element
\begin{equation}
\langle 0|J_{\mu }|T_{\mathrm{AV}}(p,\epsilon )\rangle =fm\epsilon _{\mu },
\label{eq:MElem1}
\end{equation}%
where $\epsilon _{\mu }$ is the polarization vector of the $T_{\mathrm{AV}}$
state. Then the correlation function $\Pi _{\mu \nu }^{\mathrm{Phys}}(p)$
takes the simple form
\begin{equation}
\Pi _{\mu \nu }^{\mathrm{Phys}}(p)=\frac{m^{2}f^{2}}{m^{2}-p^{2}}\left(
-g_{\mu \nu }+\frac{p_{\mu }p_{\nu }}{m^{2}}\right) +\ldots  \label{eq:CorF1}
\end{equation}%
Equation (\ref{eq:CorF1}) determines the physical or phenomenological side
of the sum rules.

The correlation function $\Pi _{\mu \nu }(p)$ calculated by employing the
quark propagators constitutes the QCD side of the sum rules. It is given by
the expression
\begin{eqnarray}
&&\Pi _{\mu \nu }^{\mathrm{OPE}}(p)=\frac{i}{4}\int d^{4}xe^{ipx}\left\{
\mathrm{Tr}\left[ \gamma ^{\alpha }\widetilde{S}^{a^{\prime }b}(-x)\gamma
^{\beta }S^{b^{\prime }a}(-x)\right] \right.  \notag \\
&&\times \mathrm{Tr}\left[ S^{ab^{\prime }}(x)\gamma _{\nu }\gamma _{\beta
}\gamma _{5}\widetilde{S}^{ba^{\prime }}(x)\gamma _{5}\gamma _{\mu }\gamma
_{\alpha }\right] -\mathrm{Tr}\left[ \gamma ^{\alpha }\widetilde{S}%
^{bb^{\prime }}(-x)\right.  \notag \\
&&\left. \times \gamma ^{\beta }S^{a^{\prime }a}(-x)\right] \mathrm{Tr}\left[
S^{ab^{\prime }}(x)\gamma _{\nu }\gamma _{\beta }\gamma _{5}\widetilde{S}%
^{ba^{\prime }}(x)\gamma _{5}\gamma _{\mu }\gamma _{\alpha }\right]  \notag
\\
&&\left. +62\ \mathrm{similar\ terms}\right\} ,  \label{eq:CF3}
\end{eqnarray}%
where $S^{ab}(x)$ is the $s$-quark propagator and
\begin{equation}
\widetilde{S}(x)=CS^{T}(x)C.  \label{eq:Prop}
\end{equation}%
In calculations we employ the $x$-space light-quark propagator
\begin{eqnarray}
&&S^{ab}(x)=i\frac{\slashed x}{2\pi ^{2}x^{4}}\delta _{ab}-\frac{m_{s}}{4\pi
^{2}x^{2}}\delta _{ab}-\frac{\langle \overline{s}s\rangle }{12}\left( 1-i%
\frac{m_{s}}{4}\slashed x\right) \delta _{ab}  \notag \\
&&-\frac{x^{2}}{192}\langle \overline{s}g_{s}\sigma Gs\rangle \left( 1-i%
\frac{m_{s}}{6}\slashed x\right) \delta _{ab}  \notag \\
&&-\frac{ig_{s}G_{ab}^{\mu \nu }}{32\pi ^{2}x^{2}}\left[ \slashed x\sigma
_{\mu \nu }+\sigma _{\mu \nu }\slashed x\right] -\frac{\slashed %
xx^{2}g_{s}^{2}}{7776}\langle \overline{s}s\rangle ^{2}\delta _{ab}  \notag
\\
&&-\frac{x^{4}\langle \overline{s}s\rangle \langle g_{s}^{2}G^{2}\rangle }{%
27648}\delta _{ab}+\frac{m_{s}g_{s}}{32\pi ^{2}}G_{ab}^{\mu \nu }\sigma
_{\mu \nu }\left[ \ln \left( \frac{-x^{2}\Lambda ^{2}}{4}\right) +2\gamma
_{E}\right]\notag \\
&&+\cdots ,  \label{eq:QProp}
\end{eqnarray}%
where $\gamma _{E}\simeq 0.577$ is the Euler constant, and $\Lambda $ is the
QCD scale parameter. We use also the notation $G_{ab}^{\mu \nu }\equiv
G_{A}^{\mu \nu }t_{ab}^{A},\ A=1,2,\ldots 8$, and $t^{A}=\lambda ^{A}/2$,
with $\lambda ^{A}$ being the Gell-Mann matrices.

The propagator (\ref{eq:QProp}) contains various light quark, gluon and
mixed condensates of different dimensions. The term $\langle \overline{s}%
g_{s}\sigma Gs\rangle $ written down in Eq.\ (\ref{eq:QProp}) as well as
other ones proportional to $\langle \overline{s}s\rangle ^{2}$, and $\langle
\overline{s}s\rangle \langle g_{s}^{2}G^{2}\rangle $ are obtained using the
factorization hypothesis of the higher dimensional condensates. It is known,
however, that the factorization assumption is not precise and violates is
the case of higher dimensional condensates  \cite{Ioffe:2005ym}. Thus, for
the condensates of dimension 10 even an order of magnitude of such a
violation is unclear. But, contributions  to sum rules arising from higher
dimensional condensates are very small, therefore, in what follows, we
ignore uncertainties generated by this violation.

At the next stage we calculate the resultant four-$x$ Fourier integrals in $%
\Pi _{\mu \nu }^{\mathrm{OPE}}(p)$. The correlation function $\Pi _{\mu \nu
}^{\mathrm{OPE}}(p)$ obtained by this way contains two Lorentz structures
which may be chosen to derive the sum rules. For our purposes terms $\sim
g_{\mu \nu }$ both in $\Pi _{\mu \nu }^{\mathrm{Phys}}(p)$ and $\Pi _{\mu
\nu }^{\mathrm{OPE}}(p)$ are convenient, because scalar particles do not
contribute to these terms. Afterwards we equate the corresponding invariant
amplitudes $\Pi ^{\mathrm{Phys}}(p^{2})$ and $\Pi ^{\mathrm{OPE}}(p^{2})$,
and find an expression in momentum space which, after some manipulations,
can be used to derive the desired sum rules. Indeed, to suppress
contributions of the higher resonances and continuum states we apply to both
sides of the obtained equality the Borel transformation. The last operation
to be carried out is continuum subtraction, which is achieved by invoking
assumption on quark-hadron duality. After these manipulations the equality
depends on auxiliary parameters of the sum rules $M^{2}$ and $s_{0}$: $M^{2}$
is the Borel parameter appeared due to corresponding transformation, $s_{0}$
is the continuum subtraction parameter that separates the ground-state and
higher resonances from each another.

To find the sum rules for $m$ and $f$ we need an additional expression which
can be obtained by acting $d/d\left( -1/M^{2}\right) $ to the first
equality. The sum rules for $m$ and $f$ have the perturbative and
nonperturbative components. The nonperturbative components contain the
quark, gluon, and mixed vacuum condensates, which appears after sandwiching
relevant terms in $\Pi ^{\mathrm{OPE}}(p)$ between vacuum states. Our
analytical results contain the nonperturbative terms up to dimension-20. We
keep all of them in numerical computations bearing in mind that higher
dimensional terms appear due to the factorization hypothesis as product of
basic condensates, and do not encompass all dimension-20 contributions.

In numerical computations we utilize the following quark and mixed
condensates: $\langle \bar{s}s\rangle =-0.8\times (0.24\pm 0.01)^{3}~\mathrm{%
GeV}^{3}$ and $\langle \overline{s}g_{s}\sigma Gs\rangle =m_{0}^{2}\langle
\bar{s}s\rangle $, where $m_{0}^{2}=(0.8\pm 0.1)~\mathrm{GeV}^{2}$. An
important ingredient of analyses is the gluon condensate $\ \langle \alpha
_{s}G^{2}/\pi \rangle =(0.012\pm 0.004)~\mathrm{GeV}^{4}$. Our sum rules
depend on the strange quark mass for which we use its value $%
m_{s}=93_{-5}^{+11}~\mathrm{MeV}$ borrowed from Ref.\ \cite%
{Tanabashi:2018oca}. \ The scale parameter $\Lambda $ can be chosen within
the limits $(0.5,\ 1)\ \mathrm{GeV}$; we utilize the central value $\Lambda
=0.75\ \mathrm{GeV}$.

A very important problem of calculations is a proper choice for the Borel $%
M^{2}$ and continuum threshold $s_{0}$ parameters. These parameters are not
arbitrary, but should meet some known requirements: At maximum of the Borel
parameter the pole contribution ($\mathrm{PC}$) has to constitute a fixed
part of the correlation function, whereas at minimum of $M^{2}$ it must be a
dominant contribution. We define $\mathrm{PC}$ in the form
\begin{equation}
\mathrm{PC}=\frac{\Pi (M^{2},s_{0})}{\Pi (M^{2},\infty )},  \label{eq:PC}
\end{equation}%
where $\Pi (M^{2},s_{0})$ is the Borel transformed and subtracted invariant
amplitude $\Pi ^{\mathrm{OPE}}(p^{2})$. The minimum of $M^{2}$ is fixed from
convergence of the sum rules, i.e. at $M_{\mathrm{min}}^{2}$ contribution of
the last term (or a sum of last few terms) cannot exceed, for example, $0.01$
part of the whole result. In the case of multi-quark hadrons at $M_{\mathrm{%
max}}^{2}$ one, as usual, requires $\mathrm{PC}>0.2$. There is an another
restriction on the lower limit $M_{\mathrm{min}}^{2}$: at $M_{\mathrm{min}%
}^{2}$ the perturbative contribution has to prevail over the nonperturbative
one.

The sum rule predictions should not depend on the parameters $M^{2}$ and $%
s_{0}$. But in real calculations $m$ and $f$ demonstrate sensitiveness to
the choice of $M^{2}$ and $s_{0}$. Hence, the parameters $M^{2}$ and $s_{0}$
have to be fixed in such a manner that to reduce this effect to a minimum.
Performed analysis allows us to find the working regions
\begin{equation}
M^{2}\in \lbrack 1.4,\ 2]~\mathrm{GeV}^{2},\ s_{0}\in \lbrack 6,\ 7]~\mathrm{%
GeV}^{2},  \label{eq:Wind1}
\end{equation}%
which obey all the aforementioned constraints.
\begin{figure}[h]
\includegraphics[width=8.8cm]{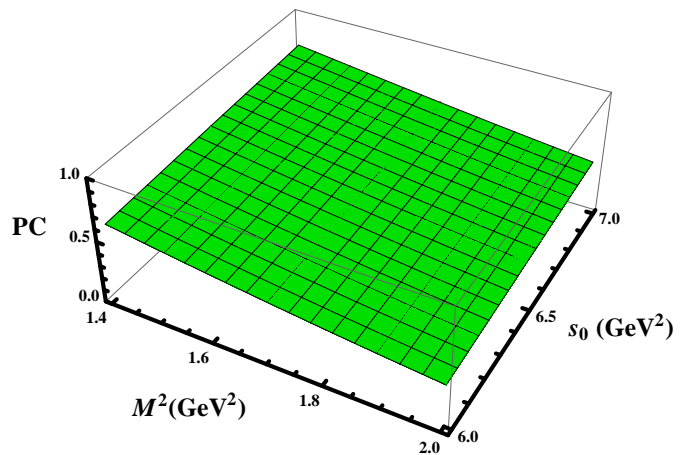}
\caption{Dependence of the pole contribution on $M^{2}$ and $s_{0}$. }
\label{fig:PC}
\end{figure}

\begin{figure}[h]
\includegraphics[width=8.8cm]{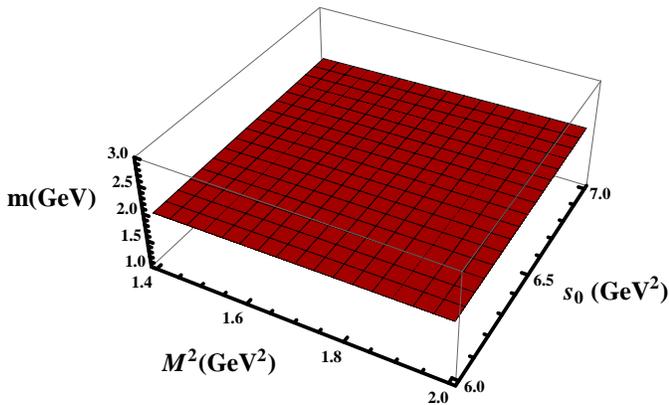}
\caption{ The mass of the tetraquark $T_{\mathrm{AV}}$ as a function of the
Borel and continuum threshold parameters.}
\label{fig:Mass1}
\end{figure}

In Fig.\ \ref{fig:PC} we depict the pole contribution as functions of $M^{2}$
and $s_{0}$: at $M^{2}=1.4$ the pole contribution is $0.68$, whereas at $%
M^{2}=2$ it becomes equal to $0.39$. The prediction for the mass $m$ is
plotted in Fig.\ \ref{fig:Mass1}, where one can see its weak dependence on
the parameters $M^{2}$ and $s_{0}$. The results for the spectroscopic
parameters of the tetraquark $T_{\mathrm{AV}}$ read:
\begin{eqnarray}
m &=&\left( 2067\pm 84\right) ~\mathrm{MeV},\   \notag \\
f &=&\left( 0.89\pm 0.11\right) \times 10^{-2}~\mathrm{GeV}^{4}.
\label{eq:Result1}
\end{eqnarray}%
Theoretical errors in the sum rule computations appear due to different
sources. The auxiliary parameters $M^{2}$ and $s_{0}$ are main sources of
these ambiguities. Errors connected with uncertainies of $m_{s}$ and vacuum
condensates are not substantial. For example, varying $m_{s}$ within tle
limits $88~\mathrm{MeV}\leq m_{s}\leq 104~\mathrm{MeV}\ $leads to
corrections $\binom{+2~}{-1}~\mathrm{MeV}$ for $m$ and $\binom{+0.0002}{%
-0.0001}$ $~\mathrm{GeV}^{4}$ for $f$. \ All of these errors are taken into
account in (\ref{eq:Result1}).

The result obtained for the mass of the axial-vector tetraquark $T_{\mathrm{%
AV}}$ is in excellent agreement with the mass of the structure $X_{1}$
reported by the BESIII Collaboration. Therefore, it is possible to identify $%
T_{\mathrm{AV}}$ with the resonance $X_{1}$. Our conclusion is also in
accord with previous theoretical predictions obtained by means of the QCD
sum rules method. Thus, the mass of the resonance $X_{1}$ was estimated in
Refs. \cite{Cui:2019roq,Wang:2019nln}
\begin{equation}
m=2000_{-90}^{+100}~\mathrm{MeV},\ m=\left( 2080\pm 120\right) ~\mathrm{MeV},
\label{eq:Previous}
\end{equation}%
respectively. As is seen, all these calculations support the assumption on
the axial-vector tetraquark nature of the structure $X_{1}$. But one needs
to explore its decay channels $X_{1}\rightarrow \phi \eta ^{\prime }$ and $%
X_{1}\rightarrow \phi \eta $, and find width of this resonance: only after
successful comparison with experimental data it is legitimate to make more
strong conclusion about $X_{1}$. We are going to address this problem in
Sec.\ \ref{sec:Decay}.


\section{ Spectroscopic parameters of the vector tetraquark $ss\overline{s}%
\overline{s}$}

\label{sec:Mass2}

In the previous section we have explored the axial-vector tetraquark $T_{%
\mathrm{AV}}$ and identified it as a candidate for the resonance $X_{1}$.
But there are two other light states which should be classified within the
four-quark picture. In the present section we are going to analyze the
vector tetraquark $T_{\mathrm{V}}=ss\overline{s}\overline{s}$ with the
quantum numbers $J^{PC}=1^{--}$ and compare the obtained result for its mass
with the experimental information of BaBar and BESIII collaborations.

\begin{widetext}

\begin{figure}[h!]
\begin{center}
\includegraphics[%
totalheight=6cm,width=8cm]{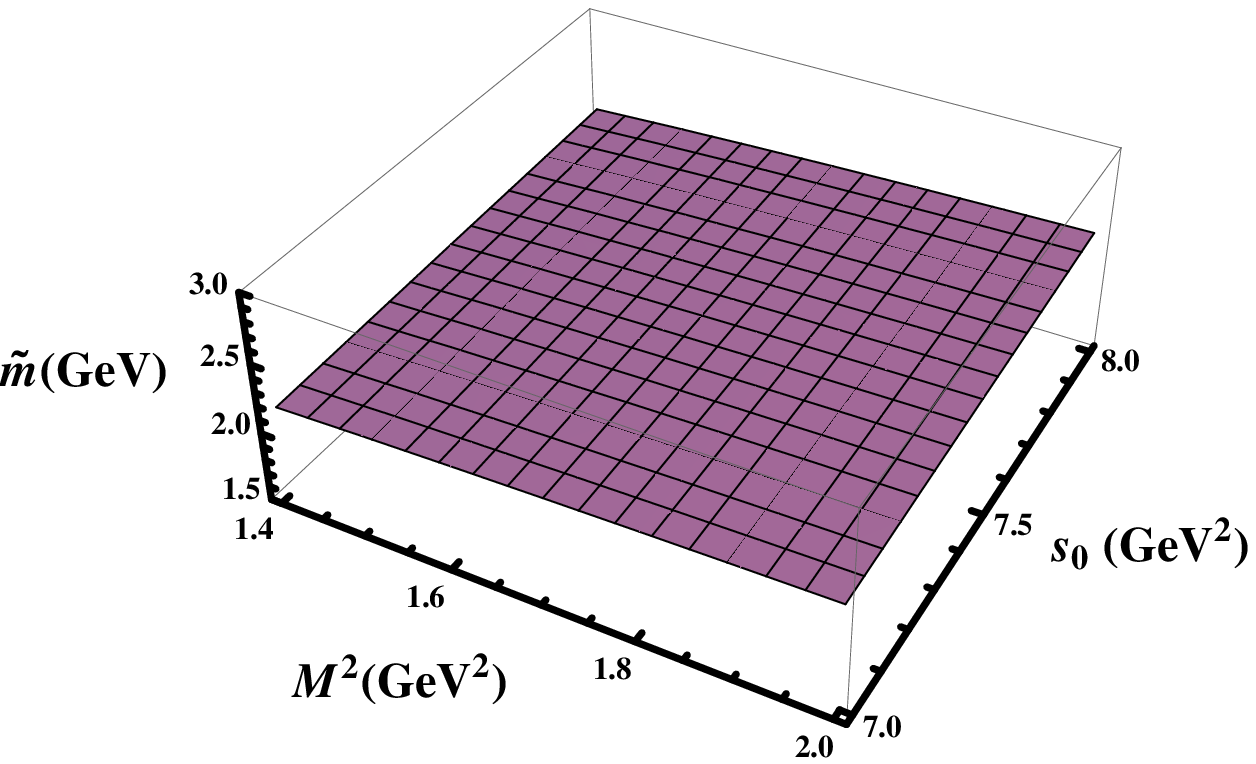}
\includegraphics[
totalheight=6cm,width=8cm]{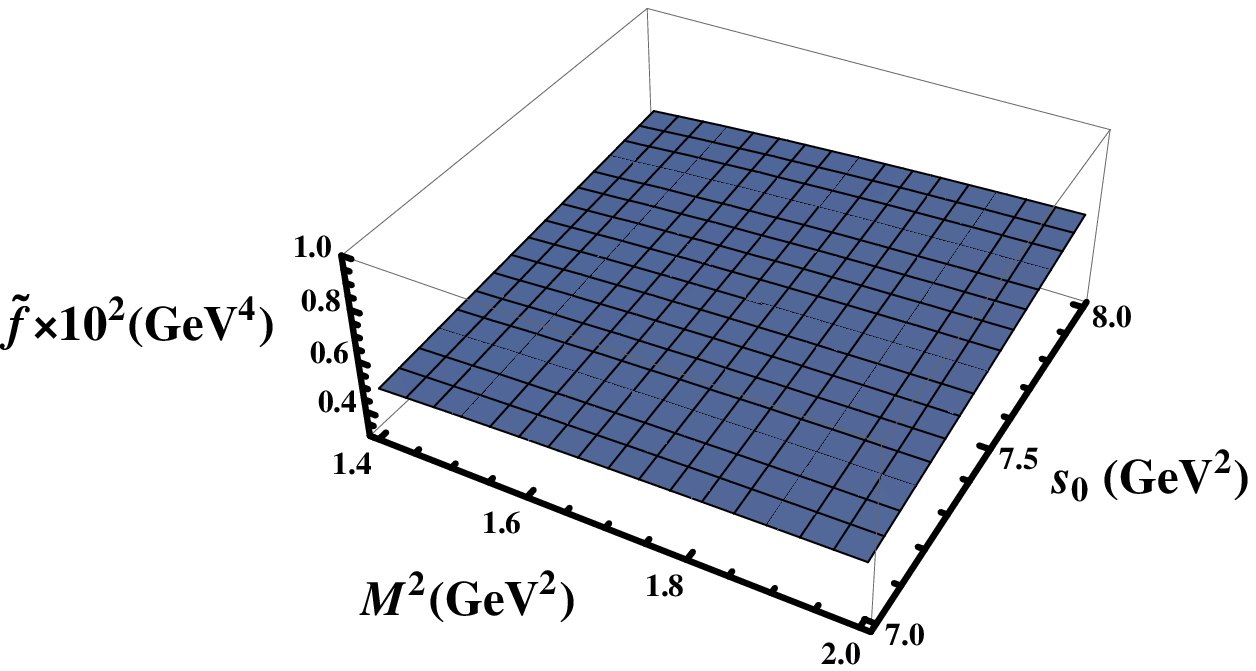}
\end{center}
\caption{ The mass (left panel) and coupling (right panel) of the vector tetraquark $T_{\mathrm{V}}$ as functions
of the Borel and continuum threshold  parameters.}
\label{fig:MassCoupl2}
\end{figure}

\end{widetext}

Calculations of the $T_{\mathrm{V}}$ tetraquark's mass $\widetilde{m}$ and
coupling $\widetilde{f}$ do not differ considerably from ones fulfilled in
the previous section. There are only some qualitative differences on which
we want to concentrate. First of all, the interpolating current for the
vector state is defined by the expression \cite{Chen:2018kuu}
\begin{eqnarray}
&&\widetilde{J}_{\mu }(x)=\left[ s_{a}^{T}(x)C\gamma _{5}s_{b}(x)\right] %
\left[ \overline{s}_{a}(x)\gamma _{\mu }\gamma _{5}C\overline{s}_{b}^{T}(x)%
\right]  \notag \\
&&-\left[ s_{a}^{T}(x)C\gamma _{\mu }\gamma _{5}s_{b}(x)\right] \left[
\overline{s}_{a}(x)\gamma _{5}C\overline{s}_{b}^{T}(x)\right] .
\label{eq:Curr2}
\end{eqnarray}%
The physical side of the sum rule is given by Eq.\ (\ref{eq:CorF1}) with
evident replacements. The correlation function $\widetilde{\Pi }_{\mu \nu }^{%
\mathrm{OPE}}(p)$ that determines the QCD side of the sum rule has the
following expression
\begin{eqnarray}
&&\widetilde{\Pi }_{\mu \nu }^{\mathrm{OPE}}(p)=i\int d^{4}xe^{ipx}\left\{
\mathrm{Tr}\left[ \gamma _{5}\widetilde{S}^{b^{\prime }b}(-x)\gamma
_{5}\gamma _{\nu }S^{a^{\prime }a}(-x)\right] \right.  \notag \\
&&\times \mathrm{Tr}\left[ S^{aa^{\prime }}(x)\gamma _{5}\widetilde{S}%
^{bb^{\prime }}(x)\gamma _{5}\gamma _{\mu }\right] -\mathrm{Tr}\left[ \gamma
_{5}\widetilde{S}^{a^{\prime }b}(-x)\gamma _{\nu }\right.  \notag \\
&&\left. \times \gamma _{5}S^{b^{\prime }a}(-x)\right] \mathrm{Tr}\left[
S^{aa^{\prime }}(x)\gamma _{5}\widetilde{S}^{bb^{\prime }}(x)\gamma
_{5}\gamma _{\mu }\right]  \notag \\
&&\left. +14\ \mathrm{similar\ terms}\right\} ,  \label{eq:CF4}
\end{eqnarray}%
The remaining operations have been explained above. Therefore we present
only final results of performed analysis. The working windows for the Borel
and continuum threshold parameters in the case of the vector tetraquark $T_{%
\mathrm{V}}$ are determined by the intervals
\begin{equation}
M^{2}\in \lbrack 1.4,\ 2]~\mathrm{GeV}^{2},\ s_{0}\in \lbrack 7,\ 8]~\mathrm{%
GeV}^{2}.  \label{eq:Wind2}
\end{equation}%
It is seen that these regions differ from ones presented in Eq.\ (\ref%
{eq:Wind1}) by only small shift of the parameter $s_{0}$. The windows (\ref%
{eq:Wind1}) comply all constraints necessary in the sum rule computations.
In fact, at $M^{2}=1.4$ the pole contribution is $60\%$, whereas at $M^{2}=2$
it is equal to $30\%$ of the whole result. Convergence of the sum rules is
also satisfied. The mass and coupling of the vector tetraquark $T_{\mathrm{V}%
}$ are:
\begin{eqnarray}
\widetilde{m} &=&\left( 2283\pm 114\right) ~\mathrm{MeV},\   \notag \\
\widetilde{f} &=&\left( 0.57\pm 0.10\right) \times 10^{-2}~\mathrm{GeV}^{4}.
\label{eq:Result2}
\end{eqnarray}%
In Fig.\ \ref{fig:MassCoupl2} we plot the spectroscopic parameters $%
\widetilde{m}$ and $\widetilde{f}$ as functions of $M^{2}$ and $s_{0}$.

Comparing the mass of the vector state $T_{\mathrm{V}}$ and experimental
information on the resonances $Y(2175)$ and $X_{2}$, one can see that it can
be identified with the $X_{2}$. In fact, difference between the masses of $%
T_{\mathrm{V}}$ and $X_{2}$ is approximately $60~\mathrm{MeV}$ smaller than
between $T_{\mathrm{V}}$ and $Y(2175)$. The similar conclusion was drawn
also in Ref.\ \cite{Lu:2019ira}. The mass $m_{X_{2}}=2227~\mathrm{MeV}$ of
the four-quark vector system $ss\overline{s}\overline{s}$ found there is
consistent with BESIII data.

The mass of the vector tetraquark $ss\overline{s}\overline{s}$ was computed
using the QCD sum rule method in Refs.\ \cite{Wang:2019nln} and \cite%
{Chen:2018kuu} as well. The prediction for the mass of this four-quark meson
$m=(3080\pm 110)~\mathrm{MeV}$ made in Refs.\ \cite{Wang:2019nln} disfavors
classifying it as the resonance $Y(2175)$. Comparing this result with recent
measurements of the BESIII\ Collaboration, we see that it also cannot be
assigned to be the resonance $X_{2}$. To study vector tetraquarks with the $%
ss\overline{s}\overline{s}$ content, in Ref. \cite{Chen:2018kuu} the authors
constructed two independent interpolating currents which couple to $%
J^{PC}=1^{--\text{ }}$states. These currents led to slightly different
predictions $m_{1}=(2410\pm 250)~\mathrm{MeV}$ and $m_{2}=(2340\pm 170)~%
\mathrm{MeV}$. In accordance with \cite{Chen:2018kuu} the first state might
correspond to a structure in the $\phi f_{0}(980)$ invariant mass spectrum
at around $2.4~\mathrm{GeV}$. The second one was interpreted in Ref. \cite%
{Chen:2018kuu} as the resonance $Y(2175)$ but, from our point of view, it is
closer to the structure $X_{2}$.

\section{Decays $X_{1}\rightarrow \protect\phi \protect\eta ^{\prime }$ and $%
X_{1}\rightarrow \protect\phi \protect\eta $}

\label{sec:Decay}

Within the framework of the QCD sum rule method the decay $X_{1}\rightarrow
\phi \eta ^{\prime }$ [and $X_{1}\rightarrow \phi \eta $] can be
investigated by means of different approaches. In fact, a key quantity to
calculate the width of this decay is the coupling $g_{X_{1}\phi \eta
^{\prime }}$ describing the strong interaction in the vertex $X_{1}\phi \eta
^{\prime }$. The coupling $g_{X_{1}\phi \eta ^{\prime }}$ can be evaluated
using, for example, the QCD three-point sum rule method. Alternatively, one
can extract it from the relevant QCD light-cone sum rule (LCSR), which has
some advantages when calculating tetraquark-meson-meson vertices containing
light mesons. The reason is that the LCSRs for tetraquark-meson-meson
vertices differ from ones involving only conventional mesons. Thus, the LCSR
for vertices of conventional mesons depends on various distribution
amplitudes (DAs) of one of the final mesons, which encode all information
about nonperturbative dynamical properties of the meson. In the case of the
tetraquark-meson-meson vertices due to four-quark nature of the tetraquark,
after contracting relevant quark fields instead of DAs of a the final meson
the sum rule contains only local matrix elements of this meson. Then to
satisfy the four-momentum conservation at vertices the momentum of a final
light meson should be set $q=0$. This leads to crucial changes in the
calculational scheme, because now one has to accompany the LCSR method with
technical tools of the soft-meson approximation \cite%
{Belyaev:1994zk,Agaev:2016dev}.

Let us consider the dominant process $X_{1}\rightarrow \phi \eta ^{\prime }$
in a detailed form. The second decay mode $X_{1}\rightarrow \phi \eta $ , as
we shall see below, can be analyzed in the same manner. The starting point
to explore the decay $X_{1}\rightarrow \phi \eta ^{\prime }$ is the
correlation function
\begin{equation}
\widehat{\Pi }_{\mu \nu }(p,q)=i\int d^{4}xe^{ipx}\langle \eta ^{\prime }(q)|%
\mathcal{T}\{J_{\mu }^{\phi }(x)J_{\nu }^{\dagger }(0)\}|0\rangle ,
\label{eq:CF5}
\end{equation}%
where $J_{\mu }^{\phi }(x)$ is the interpolating current of the $\phi $
meson
\begin{equation}
J_{\mu }^{\phi }(x)=i\overline{s}_{i}(x)\gamma _{\nu }s_{i}(x).
\label{eq:Curr3}
\end{equation}%
Following the standard recipes, we write down $\widehat{\Pi }_{\mu \nu
}(p,q) $ in terms of the physical parameters of the particles $X_{1},\ \phi $
and $\eta ^{\prime }$%
\begin{eqnarray}
&&\widehat{\Pi }_{\mu \nu }^{\mathrm{Phys}}(p,q)=\frac{\langle 0|J_{\mu
}^{\phi }(x)|\phi (p)\rangle }{p^{2}-m_{\phi }^{2}}\langle \phi (p)\eta
^{\prime }(q)|X_{1}(p^{\prime })\rangle  \notag \\
&&\times \frac{\langle X_{1}(p^{\prime })|J_{\nu }^{\dagger }|0\rangle }{%
p^{\prime 2}-m^{2}}+...,  \label{eq:CF5A}
\end{eqnarray}%
where $p^{\prime }$ and $p$, $q$ are momenta of the initial and final
particles, respectively. In Eq.\ (\ref{eq:CF5A}) contributions of excited
resonances and continuum states are indicated by dots. By utilizing the
matrix elements
\begin{eqnarray}
&&\langle 0|J_{\mu }^{\phi }(x)|\phi (p)\rangle =f_{\phi }m_{\phi
}\varepsilon _{\mu },  \notag \\
&&\langle \phi (p)\eta ^{\prime }(q)|X_{1}(p^{\prime })\rangle =g_{X_{1}\phi
\eta ^{\prime }}\left[ (p\cdot p^{\prime })(\varepsilon ^{\ast }\cdot
\varepsilon ^{\prime })\right.  \notag \\
&&\left. -(p\cdot \varepsilon ^{\prime })(p^{\prime }\cdot \varepsilon
^{\ast })\right] ,  \label{eq:Mel}
\end{eqnarray}%
one can considerably simplify $\widehat{\Pi }_{\mu \nu }^{\mathrm{Phys}%
}(p,q) $. The matrix element $\langle 0|J_{\mu }^{\phi }(x)|\phi (p)\rangle $
is expressed in terms of $\phi $ meson's mass $m_{\phi }$ , decay constant $%
f_{\phi }$ and polarization vector $\varepsilon _{\mu }$. The matrix element
of the vertex $X_{1}\phi \eta ^{\prime }$ is written down using the strong
coupling $g_{X_{1}\phi \eta ^{\prime }}$ which has to be evaluated from the
sum rule. In the soft limit $q\rightarrow 0$ we get $p^{\prime }=p$, as a
result instead of two-variable Borel transformation we have to perform
one-variable Borel transformation, which yields
\begin{eqnarray}
&&\mathcal{B}\widehat{\Pi }_{\mu \nu }^{\mathrm{Phys}}(p)=g_{X_{1}\phi \eta
^{\prime }}m_{\phi }mf_{\phi }f\frac{e^{-\overline{m}^{2}/M^{2}}}{M^{2}}
\notag \\
&&\times \left( \overline{m}^{2}g_{\mu \nu }-p_{\nu }p_{\mu }^{\prime
}\right) +\ldots ,  \label{eq:CF5B}
\end{eqnarray}%
where $\overline{m}^{2}=(m_{\phi }^{2}+m^{2})/2.$ In Eq.\ (\ref{eq:CF5B}) we
still keep $p_{\nu }\neq p_{\mu }^{\prime }$ to make clear the Lorentz
structure of the obtained expression. To derive the LCSR for the strong
coupling $g_{X_{1}\phi \eta ^{\prime }}$ we will employ the structure $\sim
g_{\mu \nu }$.

In the soft approximation the physical side of the sum rule has more
complicated structure than in the case of full LCSR method. The
complications are connected with behavior of contributions arising from
higher resonances and continuum states in the soft limit. The problem is
that in the soft limit some of these contributions even after the Borel
transformation remain unsuppressed and appear as contaminations in the
physical side \cite{Belyaev:1994zk}. Therefore, before performing the
continuum subtraction in the final sum rule they should be removed by means
of some operations. This problem is solved by acting on the physical side of
sum rule by the operator \cite{Belyaev:1994zk,Ioffe:1983ju}
\begin{equation*}
\mathcal{P}(M^{2},\overline{m}^{2})=\left( 1-M^{2}\frac{d}{dM^{2}}\right)
M^{2}e^{\overline{m}^{2}/M^{2}},
\end{equation*}%
that singles out the ground-state term. It is natural that the same operator
$\mathcal{P}(M^{2},\overline{m}^{2})$ should be applied also to the QCD side
of the sum rule. But before these manipulations the correlation function $%
\widehat{\Pi }_{\mu \nu }^{\mathrm{OPE}}(p,q)$ has to be calculated in the
soft-meson approximation and expressed in terms of the $\eta ^{\prime }$
meson's local matrix elements.

In the soft limit $\widehat{\Pi }_{\mu \nu }^{\mathrm{OPE}}(p)$ is given by
the formula
\begin{eqnarray}
&&\widehat{\Pi }_{\mu \nu }^{\mathrm{OPE}}(p)=2i\int d^{4}xe^{ipx}\left\{ %
\left[ \sigma _{\mu \rho }\gamma _{5}\widetilde{S}^{ib}(x){}\gamma _{\nu }%
\widetilde{S}^{bi}(-x)\gamma ^{\rho }\right. \right.  \notag \\
&&\left. -\gamma ^{\rho }\widetilde{S}^{ib}(x){}\gamma _{\nu }\widetilde{S}%
^{bi}(-x){}\gamma _{5}\sigma _{\mu \rho }\right] _{\alpha \beta }\langle
\eta ^{\prime }(q)|\overline{s}_{\alpha }^{a}(0)s_{\beta }^{a}(0)|0\rangle
\notag \\
&&+\left[ {}\gamma ^{\rho }\widetilde{S}^{ia}(x)\gamma _{\nu }\widetilde{S}%
^{bi}(-x)\gamma _{5}\sigma _{\mu \rho }-\gamma _{5}\sigma _{\mu \rho }%
\widetilde{S}^{ia}(x)\gamma ^{\rho }\right.  \notag \\
&&\left. \left. \times {}\widetilde{S}^{bi}(-x){}\gamma _{\nu }\right]
_{\alpha \beta }\langle \eta ^{\prime }(q)|\overline{s}_{\alpha
}^{b}(0)s_{\beta }^{a}(0)|0\rangle \right\} ,  \label{eq:CF6}
\end{eqnarray}%
where $\alpha $ and $\beta $ are the spinor indices.

It is seen that $\widehat{\Pi }_{\mu \nu }^{\mathrm{OPE}}(p)$ really depends
on local matrix elements of the $\eta ^{\prime }$ meson. But these matrix
elements should be converted to forms suitable to express them in terms of
the standard matrix elements of the $\eta ^{\prime }$ meson. To this end, we
continue calculations by employing the expansion
\begin{equation}
\overline{s}_{\alpha }^{a}s_{\beta }^{b}\rightarrow \frac{1}{12}\Gamma
_{\beta \alpha }^{j}\delta ^{ab}\left( \overline{s}\Gamma ^{j}s\right) ,
\label{eq:MatEx}
\end{equation}%
where $\Gamma ^{j}$ is the full set of Dirac matrices
\begin{equation*}
\Gamma ^{j}=\mathbf{1,\ }\gamma _{5},\ \gamma _{\lambda },\ i\gamma
_{5}\gamma _{\lambda },\ \sigma _{\lambda \rho }/\sqrt{2}.
\end{equation*}%
Then operators $\overline{s}(0)\Gamma ^{j}s(0)$ , as well as ones appeared
due to $G_{\mu \nu }$ insertions from propagators $\widetilde{S}(\pm x)$,
generate standard local matrix elements of the $\eta ^{\prime }$ meson.
Substituting Eq.\ (\ref{eq:MatEx}) into the expression of the correlation
function and carrying out the summation over color indices in accordance
with rules described in a detailed form in Ref.\ \cite{Agaev:2016dev}, we
find local matrix elements of the $\eta ^{\prime }$ meson that contribute to
$\Pi ^{\mathrm{QCD}}(p)$.

Performed analysis demonstrates that in the soft-meson approximation only
twist-3 matrix element $\langle \eta ^{\prime }|\overline{s}i\gamma
_{5}s|0\rangle $ gives non-zero contribution to the correlation function $%
\widehat{\Pi }_{\mu \nu }^{\mathrm{OPE}}(p)$. The matrix elements of the $%
\eta $ and $\eta ^{\prime }$ differ from ones of other pseudoscalar mesons:
This is connected with mixing phenomena in the $\eta -\eta ^{\prime }$
system. Thus, due to the mixing both the $\eta ^{\prime }$ and $\eta $
mesons have $\overline{s}s$ components. Of course, $\overline{s}s$ is
dominant for the $\eta ^{\prime }$ meson, whereas it plays a subdominant
role in the $\eta $ meson's quark content. Nevertheless, through the strange
components both of these mesons can appear in the final state of the decays $%
X_{1}\rightarrow \phi \eta ^{\prime }$ and $X_{1}\rightarrow \phi \eta $. \

The mixing in the $\eta -\eta ^{\prime }$ system can be described in
different basis: For our purposes, the quark-flavor basis is more convenient
than the octet-singlet basis of the flavor $SU_{f}(3)$ group. The
quark-flavor basis was used in our previous papers to study different
exclusive processes with $\eta ^{\prime }$ and $\eta $ mesons \cite%
{Agaev:2014wna,Agaev:2015faa,Agaev:2016dsg}. In the quark-flavor basis the
twist-3 matrix element $\langle \eta ^{\prime }|\overline{s}i\gamma
_{5}s|0\rangle $ can be written down in the following form%
\begin{equation}
2m_{s}\langle \eta ^{\prime }|\overline{s}i\gamma _{5}s|0\rangle =h_{\eta
^{\prime }}^{s},  \label{eq:MEl1}
\end{equation}%
where the parameter $h_{\eta ^{\prime }}^{s}$ is defined by the equality
\begin{eqnarray}
h_{\eta ^{\prime }}^{s} &=&m_{\eta ^{\prime }}^{2}f_{\eta ^{\prime
}}^{s}-A_{\eta ^{\prime }},  \notag \\
A_{\eta ^{\prime }} &=&\langle 0|\frac{\alpha _{s}}{4\pi }G_{\mu \nu }^{a}%
\widetilde{G}^{a,\mu \nu }|\eta ^{\prime }\rangle .  \label{eq:MEleta}
\end{eqnarray}%
In Eq.\ (\ref{eq:MEleta}) $m_{\eta ^{\prime }}$ and $f_{\eta ^{\prime }}^{s}$
are the mass and $s$-component of the $\eta ^{\prime }$ meson decay
constant. \ Here the $A_{\eta ^{\prime }}$ is the matrix element which
appear due to $U(1)$ axial-anomaly. The parameter $h_{\eta ^{\prime }}^{s}$
may be computed by employing Eqs.\ (\ref{eq:MEl1}) and (\ref{eq:MEleta}),
but we use its phenomenological value extracted from analysis of relevant
exclusive processes. Thus, we have
\begin{equation}
h_{\eta ^{\prime }}^{s}=h_{s}\cos \varphi ,\ h_{s}=(0.087\pm 0.006)\ \mathrm{%
GeV}^{3},
\end{equation}%
where $\varphi =39^{\circ }.3\pm 1^{\circ }.0$ is the mixing angle in the
quark-flavor basis.

Our result for the Borel transform of the invariant function $\widehat{\Pi }%
^{\mathrm{OPE}}(p^{2})$ corresponding to the structure $\sim g_{\mu \nu }$
reads%
\begin{eqnarray}
&&\widehat{\Pi }^{\mathrm{OPE}}(M^{2})=\int_{16m_{s}^{2}}^{\infty }ds\rho ^{%
\mathrm{pert.}}(s)e^{-s/M^{2}}-h_{\eta ^{\prime }}^{s}\langle \overline{s}%
s\rangle  \notag \\
&&-\langle \frac{\alpha _{s}G^{2}}{\pi }\rangle \frac{h_{\eta ^{\prime }}^{s}%
}{8m_{s}}-\frac{h_{\eta ^{\prime }}^{s}}{6M^{2}}\langle \overline{s}%
g_{s}\sigma Gs\rangle +\frac{2g_{s}^{2}h_{\eta ^{\prime }}^{s}}{81m_{s}M^{2}}%
\langle \overline{s}s\rangle ^{2},  \notag \\
&&  \label{eq:Borel2}
\end{eqnarray}%
where
\begin{equation}
\rho ^{\mathrm{pert.}}(s)=-\frac{h_{\eta ^{\prime }}^{s}}{4m_{s}\pi ^{2}}%
(s+3m_{s}^{2}).  \label{eq:Pert}
\end{equation}%
It is worth noting that the spectral density $\rho ^{\mathrm{pert.}}(s)$ is
computed as the imaginary part of the relevant term in the correlation
function. The Borel transform of nonperturbative terms are found directly
from $\widehat{\Pi }^{\mathrm{OPE}}(p^{2})$ and includes terms up to
dimension six. After acting the operator $\mathcal{P}(M^{2},\overline{m}%
^{2}) $ to $\widehat{\Pi }^{\mathrm{OPE}}(M^{2})$ one can perform the
continuum subtraction. This implies replacement $\infty \rightarrow s_{0}$
in the first term, whereas terms $\sim (M^{2})^{0}$ and $\sim 1/M^{2}$
should be left in their original forms \cite{Belyaev:1994zk}.

The width of the decay $X_{1}\rightarrow \phi \eta ^{\prime }$ is determined
by the formula%
\begin{equation}
\Gamma (X_{1}\rightarrow \phi \eta ^{\prime })=\frac{g_{X_{1}\phi \eta
^{\prime }}^{2}m_{\phi }^{2}}{24\pi }|\overrightarrow{p}|\left( 3+\frac{2|%
\overrightarrow{p}|^{2}}{m_{\phi }^{2}}\right) ,  \label{eq:DWformula}
\end{equation}%
where
\begin{eqnarray}
&&|\overrightarrow{p}|=\frac{1}{2m}\left( m^{4}+m_{\phi }^{4}+m_{\eta
^{\prime }}^{4}-2m^{2}m_{\phi }^{2}\right.  \notag \\
&&\left. -2m^{2}m_{\eta ^{\prime }}^{2}-2m_{\phi }^{2}m_{\eta ^{\prime
}}^{2}\right) ^{1/2}.  \label{eq:PP}
\end{eqnarray}%
In numerical computations, the parameters $M^{2}$ and $s_{0}$ are varied
within the limits
\begin{equation}
M^{2}\in \lbrack 1.4,\ 2]~\mathrm{GeV}^{2},\ s_{0}\in \lbrack 6.2,\ 7.2]~%
\mathrm{GeV}^{2}.  \label{eq:Wind3}
\end{equation}%
The mass of the final-state mesons $\phi $ and $\eta ^{\prime }$ are
borrowed from Ref. \cite{Tanabashi:2018oca}
\begin{eqnarray}
m_{\phi } &=&(1019.461\pm 0.019)~\mathrm{MeV},  \notag \\
m_{\eta ^{\prime }} &=&(957.78\pm 0.06)~\mathrm{MeV},  \notag \\
f_{\phi } &=&(215\pm 5)~\mathrm{MeV.}  \label{eq:Finalmesons}
\end{eqnarray}%
Calculations lead to the following results:%
\begin{eqnarray}
&&g_{X_{1}\phi \eta ^{\prime }}=(2.82\pm 0.54)~\mathrm{GeV}^{-1},  \notag \\
&&\Gamma (X_{1}\rightarrow \phi \eta ^{\prime })=(105.3\pm 28.6)~\mathrm{MeV}%
.  \label{eq:DW1}
\end{eqnarray}

The $X_{1}\rightarrow \phi \eta ^{\prime }$ is the main $S$-wave decay
channel of the tetraquark $X_{1}$. The partial width of the second process $%
X_{1}\rightarrow \phi \eta $ can be easily evaluated by employing
expressions obtained in the present section. The differences between two
decays stem from the twist-3 matrix element, which for this decay is given
by the formula
\begin{equation}
2m_{s}\langle \eta |\overline{s}i\gamma _{5}s|0\rangle =-h_{s}\sin \varphi ,
\end{equation}%
and from the $\eta $ meson mass $m_{\eta }=(547.862\pm 0.018)~\mathrm{MeV}$
[see, Eq.\ (\ref{eq:PP}) ]. Computations yield the following predictions%
\begin{eqnarray}
|g_{X_{1}\phi \eta }| &=&(0.85\pm 0.22)~\mathrm{GeV}^{-1},  \notag \\
\Gamma (X_{1} &\rightarrow &\phi \eta )=(24.9\pm 9.5)~\mathrm{MeV}.
\label{eq:DW2}
\end{eqnarray}%
Let us note that $|g_{X_{1}\phi \eta }|$ has been extracted from the sum
rule at $s_{0}\in \lbrack 5.8,\ 6.8]~\mathrm{GeV}^{2}.$

Saturating the full width of the $X_{1}$ resonance by these two decays we
get:%
\begin{equation}
\Gamma =(130.2\pm 30.1)~\mathrm{MeV}.  \label{eq:FullW}
\end{equation}%
This estimate does not coincide with full width of the resonance $X_{1}$,
but is comparable with it.


\section{Summary and conclusions}

\label{sec:Conc}

In the present work we have studied the axial-vector and vector tetraquarks
with the quark content $ss\overline{s}\overline{s}$. The mass $m=\left(
2067\pm 84\right) ~\mathrm{MeV}$ of the axial-vector state obtained in the
present work is in excellent agreement with measurements of the BESIII
Collaboration. The width of this state $\Gamma =\left( 130.2\pm 30.1\right) ~%
\mathrm{MeV}$ within both theoretical and experimental errors is consistent
with the data. These facts have allowed us to interpret the resonance $%
X(2100)$ discovered recently the BESIII Collaboration as an axial-vector
state with quark content $ss\overline{s}\overline{s}$.

The vector $ss\overline{s}\overline{s}$ tetraquark with the mass $\widetilde{%
m}=\left( 2283\pm 114\right) ~\mathrm{MeV}$ can be identified with the
structure $X(2239)$ rather than with the resonance $Y(2175)$. There is still
the light resonance $Y(2175)$ which in the present scheme may be considered
as a conventional vector meson, because its mass is small to classify it as
a vector tetraquark. One should take into account also a possible structure
in the $\phi f_{0}(980)$ invariant mass spectrum at $2.4~\mathrm{GeV}$. In
our present work we have tried to answer questions on nature of two light
resonances. It is evident that the whole family of such structures deserves
further detailed investigations.

\end{document}